\documentclass[11pt,a4paper]{article}

\usepackage[english]{babel}

\pdfoutput=1 

\RequirePackage{ifpdf}
\usepackage{amsmath}
\usepackage{mathtools}

\usepackage{jheppub}
\usepackage{pstricks}
\usepackage[final]{pdfpages}
\usepackage{ifpdf}
\usepackage{slashed}
\usepackage{subfiles}

\usepackage{hyperref}

\usepackage{color}
\usepackage{graphics}
\usepackage{multirow}
\usepackage{etoolbox} 
\usepackage{fixmath}

\usepackage{notoccite}

\usepackage{amsfonts}
\usepackage{float}

\usepackage[normalem]{ulem}
\usepackage{tabularx,siunitx}

\newcommand{\ocean}{{\tt OCEANN}}
\newcommand{\abiss}{{\tt Abiss}}
\def\seasyde{{\tt SeaSyde}}
\def\amflow{{\tt AMFlow}}
\newcommand{\kira}{{\tt KIRA}}
\newcommand{\oaa}{${\cal O}(\alpha^2)~$}
\def\M{{\cal M}}
\def\unM{\hat{\cal M}}

\def\asrhat{\left( \frac{\hat\alpha_s}{4 \pi} \right)}
\def\aem{\left( \frac{\alpha}{4 \pi} \right)}
\def\I{{\cal I}}

\title{Towards the two-loop electroweak corrections to the  Drell-Yan process: the infrared structure}

\author[a,b]{Tommaso Armadillo,}
\author[c]{Simone Devoto,}
\author[b]{Michele Dradi,}
\author[b,d]{Alessandro Vicini}

\affiliation[a]{Centre for Cosmology, Particle Physics and Phenomenology (CP3), Université catholique de Louvain, Chemin du Cyclotron, 2, B-1348 Louvain-la-Neuve, Belgium}
\affiliation[b]{Dipartimento di Fisica ``Aldo Pontremoli'',
  University of Milano and INFN, Sezione di Milano, I-20133 Milano, Italy}
\affiliation[c]{Department of Physics and Astronomy, Ghent University, 9000 Ghent, Belgium}
\affiliation[d]{Tsung-Dao Lee Institute, Shanghai Jiao Tong University, Shanghai, 200120, China}

\emailAdd{tommaso.armadillo@uclouvain.be}
\emailAdd{simone.devoto@ugent.be}
\emailAdd{michele.dradi@mi.infn.it}
\emailAdd{alessandro.vicini@mi.infn.it}

\abstract{
We discuss  the lepton-pair production process in Quantum Electrodynamics.
We present the  ultraviolet-renormalised and infrared-subtracted finite contribution of the second-order virtual corrections to the inclusive lepton-pair production cross section $u\bar u\to e^+e^-$.
The results are obtained within a new computational framework, \ocean, 
developed in view of the evaluation of the exact two-loop electroweak virtual corrections to high-energy scattering processes.
One of the key methodological features in this approach is the representation of the scattering amplitude with arbitrary precision at every stage of the calculation.
The analysis in QED allows us to address the treatment of the infrared structure of the process and it can be easily extended to the complete electroweak Standard Model case.
The perfect agreement 
with the literature, 
for this subset of corrections, provides a non trivial validation of the general framework.
}

\preprint{TIF-UNIMI-2025-22}

\begin{document}
\allowdisplaybreaks[4]
\unitlength1cm
\maketitle
\flushbottom

\section{Introduction}
\setcounter{equation}{0}
\label{sec:intro}
The hadro-production of lepton-antilepton or lepton-antineutrino pairs, known respectively as neutral current (NC) and charged current (CC) Drell-Yan (DY) processes \cite{Drell:1970wh}, is crucial at hadron colliders since it provides the environment for a precise study of the gauge sector of the Standard Model (SM) of the electroweak (EW) and strong interactions. In particular, the NC DY process is important for the determination of the $Z$-boson mass and the effective weak mixing angle, and the CC DY process is important for the determination of the $W$-boson mass. These three EW parameters are known at the moment with small relative errors, 0.01\% for the boson masses and 0.1\% for the effective weak mixing angle \cite{Group:2012gb,Aaboud:2017svj,Aaltonen:2018dxj,ATLAS:2018gqq,CMS-PAS-SMP-22-010}, estimated within a fitting procedure of the measured kinematical distributions of the final state leptons with corresponding theoretical templates \cite{CarloniCalame:2016ouw,Chiesa:2019nqb,Bagnaschi:2019mzi,Behring:2021adr,Rottoli:2023xdc,Torrielli:2023tiz}.
The searches for new heavy particles, beyond those predicted by the SM, can be performed with the DY observables, exploring the few TeV regime. The prospects in the High-Luminosity phase of the LHC (HL-LHC) are extremely promising, with a statistical error below the 1\% level for invariant masses of about 1 TeV, which will allow a test of the theory predictions at the quantum level.
Due to the high precision of the experimental data, any statistically meaningful comparison requires correspondingly precise  theoretical predictions, which must include higher-order perturbative corrections.

The region of large invariant mass of the lepton pair is well understood from the point of view of the strong interactions, after the completion of the calculation of the third order corrections in Quantum Chromodynamics (QCD) \cite{Duhr:2020seh,Duhr:2020sdp,Duhr:2021vwj}.
The large size of the Next-To-Leading Order (NLO) EW \cite{Dittmaier:2001ay,Baur:2004ig,Zykunov:2006yb,Arbuzov:2005dd,CarloniCalame:2006zq,Baur:2001ze,Zykunov:2005tc,CarloniCalame:2007cd,Arbuzov:2007db,Dittmaier:2009cr} corrections, and in turn of the Next-To-Next-To-Leading Order (NNLO) mixed QCD-EW corrections \cite{Bonciani:2021zzf,Buccioni:2022kgy,Armadillo:2024nwk}, suggests that also the NNLO EW effects might potentially be large, at the several percent level, as it appears for instance in \cite{Jantzen:2005az,Denner:2006jr}. If confirmed, effects of this size could not be neglected for a meaningful comparison with the experimental data.
In a complementary perspective, the residual uncertainties of a NLO-EW calculation of the NC DY process, due to renormalisation ambiguities, range between 1 and 10\% in the TeV region, with a strong limitation of the discovery potential of any signal of physics beyond the SM \cite{Chiesa:2024qzd}, based on current tools.
The NNLO-EW corrections will also be relevant for precision phenomenology at HL-LHC in the 80-400 GeV interval, where sub-percent errors are expected. 
All these remarks motivate a thorough investigation of this set of quantum effects, which represents, on the other hand, a major theoretical and computational challenge.

The DY process has always represented, in the study of the QCD and EW interactions, a benchmark: for its physical relevance and also for the challenges in the development of new computational techniques  of higher-order quantum corrections.
A major progress has been witnessed in QCD, spanning four decades, with the development of new mathematical tools to compute the Feynman integrals and to evaluate the scattering amplitudes for the production of a colorless final state, involving massless partons up to third perturbative order, both at inclusive and differential level, possibly with the inclusion, beyond the fixed-order results, of the resummation of logarithmically enhanced terms to all orders in the coupling constant expansion
\cite{Altarelli:1979ub,Hamberg:1990np,Harlander:2002wh,Duhr:2020seh,Duhr:2020sdp,Duhr:2021vwj,Anastasiou:2003yy,Anastasiou:2003ds,Melnikov:2006kv,Catani:2009sm,Catani:2010en,Camarda:2021ict,Chen:2022cgv,Neumann:2022lft,Campbell:2023lcy,Chen:2021vtu,Chen:2022lwc,Moch:2005ky,Laenen:2005uz,Ravindran:2005vv,Ravindran:2006cg,deFlorian:2012za,Ahmed:2014cla,Catani:2014uta,Li:2014afw,Ajjath:2020ulr}.
The progress for the first order EW corrections is comparatively more recent: while the one-loop Feynman integrals are well established \cite{tHooft:1978jhc}, the development of several tools for the automated generation of the scattering amplitudes \cite{Buccioni:2019sur,Actis:2016mpe,Denner:2017wsf} allows to study arbitrary processes, despite the large size of the analytical expression of the amplitude, the latter due to the presence of several different masses of the gauge and Higgs bosons and of the fermions.
The combination of the difficulties in the amplitude generation and in the evaluation of the Feynman integrals relevant for the NNLO mixed QCD-EW corrections have been faced in steps of increasing complexity, first considering the combination of first order QCD and EW corrections in Monte Carlo simulation tools \cite{Balossini:2008cs,Balossini:2009sa, Bernaciak:2012hj,Barze:2012tt,Barze:2013fru,Frederix:2018nkq,Chiesa:2024qzd}, then studying in detail the resonant production of a gauge boson \cite{Delto:2019ewv,Cieri:2020ikq,Bonciani:2016wya,Bonciani:2019nuy,Bonciani:2020tvf,Bonciani:2021iis,Buccioni:2020cfi,Behring:2020cqi,Dittmaier:2014qza,Dittmaier:2015rxo,Dittmaier:2020vra,Dittmaier:2024row}, and eventually with the discussion of the complete amplitude for the production of a lepton pair \cite{deFlorian:2018wcj,Bonciani:2021zzf,Buccioni:2022kgy,Armadillo:2022bgm,Heller:2020owb,Heller:2019gkq,Hasan:2020vwn,Buonocore:2021rxx,Armadillo:2024nwk}. New techniques based on the series expansion solution of the differential equations satisfied by the so called Master Integrals (MIs) have opened the way to the evaluation of a general Feynman integral with an arbitrary set of internal complex valued masses \cite{Moriello:2019yhu,Hidding:2020ytt,Armadillo:2022ugh,Liu:2022chg,Prisco:2025wqs,Armadillo:2025mvu}.

The evaluation of the NNLO EW corrections to the cross section $\sigma(pp\to \ell^+\ell^-+X)$ is a formidable challenge, which deserves the development of a dedicated computational framework, where all the different conceptual and computational issues relevant in the calculation are addressed in a fully consistent way. We introduce here \ocean\,(Organised Calculation of Electroweak Amplitudes at NNLO), a framework that combines the usage of different existing tools and allows an experienced user to perform a full two-loop EW calculation in a systematic and sustainable way.
The challenge is evident in the evaluation of all the contributions to the inclusive cross section, because of the high target precision in the evaluation of the subprocesses where the lepton pair is accompanied by real radiation, but also and foremost because of the complexity of the two-loop virtual corrections to the lowest order partonic subprocesses.
Several distinct physics items have to be discussed for the latter: the ultraviolet (UV) renormalisation in the presence of unstable particles and the description of the gauge boson resonances; the evaluation of the Feynman integrals with internal complex-valued masses; the corrections due to an internal closed fermionic loop and the study of the couplings of the gauge bosons to fermions; the subtraction of all the infrared (IR) divergences due to the emission of photons; the identification of the corrections due to massive weak bosons, which have a soft logarithmic behavior in the high-energy limit.
Each of these items raises specific mathematical and computational issues, which deserve a detailed discussion.

The main methodological choice that defines the \ocean\, framework is that all the elements contributing to the expression of the scattering amplitude are evaluated with arbitrary numerical precision, an approach ideally suited for the calculation of higher-order EW corrections.
In the QCD literature, the cancellation of the divergent terms, which appear in the UV renormalisation and in the subtraction of the IR divergences, is typically achieved at symbolical level, with the coefficients written as exact numbers.
However, this is not necessarily viable at the two-loop QCD-EW or purely EW level, because of the presence of new kinds of integrals, whose solution is not known in closed form. 
Nevertheless, the series expansion approach implemented within \ocean\, allows for an evaluation of such integrals with arbitrary numerical precision. This feature guarantees the exactness of the cancellation of the divergent terms, also in the most difficult cases. The same approach naturally applies to all the integrals present in a full two-loop EW calculation, including those whose solution is still unknown in the literature.
Furthermore, embedding the calculation in a framework leads to a uniform treatment and bookkeeping of all the integrals which appear in the scattering amplitude, a feature which is particularly convenient in view of an advanced automation of the calculation and deployment on large computing clusters.

In this paper we focus on the IR structure of the virtual amplitudes.
The universality of the IR divergences (cfr. \cite{Agarwal:2021ais} and references therein) allows a complete explicit discussion of this topic in a slightly simplified framework: we present here the two-loop virtual corrections to the production of a massive lepton pair in quark-antiquark annihilation, and for the sake of definiteness we consider $u\bar u\to e^+e^-$, in QED, including the UV renormalisation and the subtraction of the IR divergences.
The discussion of the NNLO QED corrections in QED to the scattering of 2 fermions into 2 fermions is mature in the literature
\cite{WorkingGrouponRadiativeCorrections:2010bjp,CarloniCalame:2011aa,CarloniCalame:2011zq,Banerjee:2020rww,CarloniCalame:2020yoz,Budassi:2021twh,Aliberti:2024fpq}
and a renovated interest has lead to a closed form representation of the two-loop amplitudes
\cite{Bonciani:2021okt,Delto:2023kqv}. The latter results are an important benchmark for the validation of our computational framework and to demonstrate its readiness to the evaluation of the two-loop EW corrections.

The paper is organised in the following way.
In Section \ref{sec:infra}, we describe the general properties of the IR divergences in high-energy scattering processes and our subtraction formalism.
In Section \ref{sec:process} we detail our case study, namely the evaluation of the two-loop virtual amplitude for the process $u\bar u\to e^+e^-$ in QED, including some details of the UV renormalisation.
In Section \ref{sec:results} we comment on the workflow implemented in \ocean\, to increase the level of automation in the evaluation of two-loop EW corrections, we give an explicit example with the two-loop virtual corrections in QED, and we demonstrate the numerical control on the cancellation of UV and IR divergences.
Finally, in Section \ref{sec:conclusions} we draw our conclusions.

\section{Infrared Singularities and Universal Pole Structure}
\label{sec:infra}
The bare amplitude $\hat {\mathcal M}$ for an arbitrary scattering process can be expressed in terms of a series expansion in the unrenormalised strong ($\hat \alpha_S$) and electroweak ($\hat \alpha$) coupling constants,
\begin{equation}
\label{eq:expansion_unM}
    |\unM \rangle =  4\pi\hat\alpha \sum_{i,j} \asrhat^i \hat\aem^j |\unM^{(i,j)}\rangle\;.
\end{equation}
Despite the full generality of Equation~(\ref{eq:expansion_unM}), the IR divergent structure of the bare amplitude $\hat {\mathcal M}$ can be captured in a process-independent way by exploiting the universal structure of IR divergences~\cite{Agarwal:2021ais}.
This is made possible, after the UV renormalisation\footnote{We use symbols without a hat to describe the UV renormalised quantities.}, by the introduction of suitable subtraction operators $\I^{(i,j)}$, which allow to write at each perturbative order the finite contribution of the amplitude, $|\M^{(i,j),fin}\rangle$, with the follow iterative definition, where $| \M^{(0,0)} \rangle$ is the lowest-order amplitude:
\begin{align}
      &| \M^{(0,0),fin} \rangle = | \M^{(0,0)} \rangle\;,\\
      \label{eq:subNNLO}
      &| \M^{(i,j),fin} \rangle = | \M^{(i,j)} \rangle - \sum_{k=0}^i \sum_{l=0}^j \;{\I}^{(k,l)} \;| \M^{(i-k,j-l),fin} \rangle\;.
\end{align}
The operators $\I^{(i,j)}$ describe the emission of massless gauge bosons such as gluons and photons and the production of additional light fermion pairs.
The scattering amplitude diverges when the emitted particles have vanishing energy (soft limit) or form a vanishing angle with their emitter (collinear limit). The integration over the real phase space of the emitted particles, or in the case of virtual corrections the integration over the loop momentum, are divergent and we regularise them working in $d=4-2\epsilon$ space-time dimensions.  The divergences appear as poles in $\epsilon$.  In the case of emissions off the massive final state, the emitter mass $m$ acts as a natural regulator, the collinear divergence represented by a $1/\epsilon$ pole is absent, but in the amplitude we find logarithms of $m$. Dimensional regularization is adopted also to discuss the UV renormalisation.
In the case of EW corrections, the conservation of the vector current implies that the impact of genuinely weak corrections exactly vanishes in the divergent emission factors, whose expression receives only QED 
contributions.
The universality of the IR divergent factors implies that their expression depends only on the identity of the external particles of the scattering under study, but not on the details of the hard interaction; in other words, the discussion of the IR structure of the $u\bar u\to e^+e^-$ process is the same, irrespective of the fact that we have a QED or an EW interaction driving the scattering process.

Considering the case of QED corrections up to second order, Equation~(\ref{eq:subNNLO}) takes the form
\begin{align}
  | \M^{(0,1),fin} \rangle &= 
  | \M^{(0,1)} \rangle - \;\I^{(0,1)} | \M^{(0,0)} \rangle
\\
  | \M^{(0,2),fin} \rangle &= 
  | \M^{(0,2)} \rangle - \; \I^{(0,2)} | \M^{(0,0)} \rangle
                      -  \;{\I}^{(0,1)} | \M^{(0,1),fin} \rangle\,.
 \label{eq:subtracted}
\end{align} 
We restrict the discussion to the pure QED corrections to a scattering process $2\to n$, where the final state only contains massive emitters. In this case, the infrared structure of the amplitude is encapsulated in the subtraction operators $\I^{(0,1)}$ and $\I^{(0,2)}$, defined as
\begin{align}
\label{eq:I01}
\frac 1{S_\epsilon}\,\I^{(0,1)} =& \
\frac{\Gamma'_0}{4\epsilon^2}+\frac{\Gamma_0}{2\epsilon}
\,,\\
\label{eq:I02}
\frac 1{S_\epsilon^{2}}\,\I^{(0,2)} =& \ \frac{(\Gamma'_0)^2}{32\epsilon^4}+\frac{\Gamma'_0}{8\epsilon^3}\left(\Gamma_0-\frac 32 \beta_0\right)+\frac{\Gamma_0}{8\epsilon^2}\left(\Gamma_0-2\beta_0\right)+\frac{\Gamma'_1}{16\epsilon^2}+\frac{\Gamma_1}{4\epsilon}
-\frac23 \sum_{k}^{\text{massive}} n_k q_k^2 
\\&\nonumber
\times
\left\{\frac1{2\epsilon^2} \Gamma'_0\log\left(\frac{\mu_R^2}{m_k^2}\right) + \frac1{4\epsilon}\left[ \Gamma'_0\left(\log^2\left(\frac{\mu_R^2}{m_k^2}\right)+\frac{\pi^2}6 \right)+4\Gamma_0 \log\left(\frac{\mu_R^2}{m_k^2}\right)\right]\right\}\;,
\end{align}
where $S_\epsilon=(4\pi e^{-\gamma_E})^\epsilon$ and $\mu_R$ is the renormalisation scale. 
The structure of the subtraction operators is governed by the functions $\Gamma$ and $\Gamma'$, which can be obtained from the abelianisation of the QCD soft and cusp anomalous dimensions~\cite{Becher:2009kw, Korchemsky:1987wg}
\begin{align}
    \label{eq:gamma}
     \Gamma=& \
     \sum_{i} \gamma_l^i +Q_{1} Q_{2} \gamma^{\text{cusp}}(\alpha)\log\left(\frac{\mu_R^2}{- 2p_{1}\cdot p_{2} }\right)
     \\\nonumber&
     +\sum_{j} \gamma_h^j -\sum_{j_1 \neq j_2} \frac{Q_{j_1} Q_{j_2}}2 \gamma^{\text{cusp}}(\alpha) \beta_{j_1, j_2}\coth\beta_{j_1, j_2}
     \\\nonumber&
     - \sum_{i,j} Q_i Q_j \gamma^{\text{cusp}}(\alpha)\log\left(\frac{\mu_R\, m_j}{ 2 p_i\cdot p_j}\right)\;,
     \\\label{eq:gamma'}
     \Gamma'=&-\sum_{i} Q_i^2 \gamma^{\text{cusp}}(\alpha)\;,
     \\\label{eq:gammacusp1}
     \gamma^{\text{cusp}}(\alpha)=&\ \frac{\alpha}{4\pi}\left[4-\frac{80}{9} \,\left(\frac{\alpha}{4\pi}\right) \sum_{k}^{\text{massless}} n_k q_k^2 + \mathcal O(\alpha^3)\right]\; \;.
\end{align}
We refer to massless initial state particles, with momentum $p_i$ and electric charge $Q_i$, while with the labels $j=3, \dots, n$ we refer to the massive final state emitters, with mass $m_j$, momentum $p_j$ and electric charge $Q_j$. 
The first line of Equation~(\ref{eq:gamma}) thus represents the contribution coming from initial-state radiation, with $\gamma_l^i$ defined as
\begin{align}
    \gamma_l^i=\frac{\alpha}{4\pi} Q_i^2\left\{
    -3
    +\left(\frac{\alpha}{4\pi}\right) 
    \left[
    Q_i^2 \left(-\frac 32 + 2\pi^2-24\zeta_3\right)
    +\left(\frac{130}{27}+\frac{2\pi^2}3\right)\sum_{k}^{\text{massless}} n_k q_k^2
    \right]
    +\mathcal O(\alpha^3)\;\right\}.
    \label{eq:massless}
\end{align}
In Equation~(\ref{eq:massless}) the sum runs over the particles considered massless in the computation, with multiplicity $n_k$ (3 for the coloured quarks, 1 for the leptons) and electric charge $q_k$, giving the contribution of the soft and collinear emission of light fermion pairs from the initial-state partons.
The second line of Equation~(\ref{eq:gamma}) instead contains the contribution coming from the final-state particles, with $\gamma_h^j$ defined as 
\begin{align}
    \gamma_h^j=\frac{\alpha}{4\pi} Q_j^2\left[-2
    +\left(\frac{\alpha}{4\pi}\right)\frac{40}{9} \sum_{k}^{\text{massless}} n_k q_k^2
    +\mathcal O(\alpha^3)\right]\;.
\end{align}
Finally, the third line of Equation~(\ref{eq:gamma}) contains the interference between initial-state and final-state emissions.
We label with $\beta_{j_1, j_2}$ the function defined by the relation
\begin{equation}
    \cosh \beta_{j_1, j_2}= -\frac{p_{j_1}\cdot p_{j_2}}{ m_{j_1} m_{j_2}}\;.
\end{equation}
The physically allowed values for $\cosh \beta_{j_1, j_2}$ with $j_1$ and $j_2$ outgoing are $\cosh \beta_{j_1, j_2} \leq 1$, corresponding to $\beta_{j_1, j_2}=-b +i\pi$ with real $b\geq0$.

The symbols $\Gamma_n$, $\Gamma'_n$ in Equation~(\ref{eq:I01}, \ref{eq:I02}) indicate the coefficients of the series expansion in the weak coupling $\alpha$ of the corresponding anomalous dimensions
\begin{align}
    \Gamma^{(\prime)}= \ \sum_{n\geq 0} \left(\frac{\alpha}{4\pi}\right)^{n+1}\Gamma_n^{(\prime)}\;.
\end{align}
The sum in Equation~(\ref{eq:I02}) takes care of the collinear emission of massive fermion pairs and of soft emission with a massive quark loop. As such, it runs over all the particles $k$ with electric charge $q_k$ considered with a non-zero mass $m_k$ in the calculation.
   
The expression of ${\cal I}^{(0,2)}$ receives a contribution from the renormalisation of the one-photon emission vertex, which has been performed in the $\overline{MS}$ scheme, and contributes with a $\beta_0$ term
\begin{align}
     \label{eq:beta0}
     \beta_0=&-\frac{4}{3} \, \sum_{k}^{\text{massless}} n_k \, q_k^2 \;.
\end{align} 
The expressions in Eqs.~(\ref{eq:I01}, \ref{eq:I02}) are fully consistent with those presented in \cite{Bonciani:2021okt}, and have been derived with the abelianization of the QCD subtraction operator presented in \cite{Becher:2009cu,Becher:2009kw, Ferroglia:2009ii}.

\section{Two-loop corrections to lepton-pair production in QED}
\label{sec:process}
\subsection{The process}
The process under study is the production of a  pair of massive leptons in quark-antiquark annihilation, for the sake of definiteness an up-quark pair annihilating into an electron pair 
\begin{equation}
 u(p_1) + \bar{u}(p_2) \rightarrow e^+(p_3) + e^{-} (p_4) \,.
\label{eq:process}
\end{equation}
The electron mass is labeled by $m_e$. 
The Mandelstam variables are defined as
\begin{equation}
 s = (p_1+p_2)^2, \,\  t = (p_1-p_3)^2, \, \ u = (p_2-p_3)^2 \,\,\, {\rm with} \,\, s+t+u= 2 m_e^2 \,,
\end{equation}
while the on-shell conditions of the external particles are
\begin{equation}
 p_1^2 = p_2^2 =  0; ~ ~ ~ p_3^2 = p_4^2  = m_e^2.
\end{equation}
In this paper we plan to study the IR structure of the second order purely EW corrections to this process, and to achieve this goal we limit the discussion to the evaluation in QED of the interference terms\footnote{We do not discuss the interference term $\langle \unM^{(0,1)} | \unM^{(0,1)} \rangle $, also contributing at \oaa, which can be easily obtained with any NLO amplitude generator.}
\begin{equation}
  \langle \unM^{(0,0)} | \unM^{(0,1)} \rangle \,, ~~
  \langle \unM^{(0,0)} | \unM^{(0,2)} \rangle \,,
  \label{eq:interferences}
\end{equation}
which contribute  to the corrections to the unpolarised squared matrix element of the process in Equation~\eqref{eq:process}, at \oaa\!\!.
The evaluation of the finite part of the two-loop virtual amplitude entails first the discussion of the UV renormalisation, followed by the subtraction of the IR divergences.
\begin{figure}
\begin{minipage}{0.3\textwidth}
\centering
\includegraphics[width=0.95\textwidth]{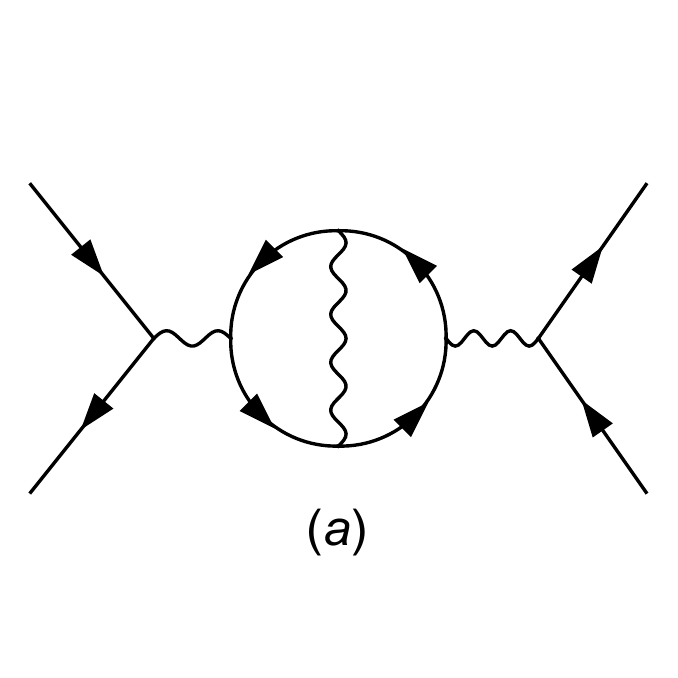}
\end{minipage}
\begin{minipage}{0.27\textwidth}
\centering
\includegraphics[width=0.98\textwidth]{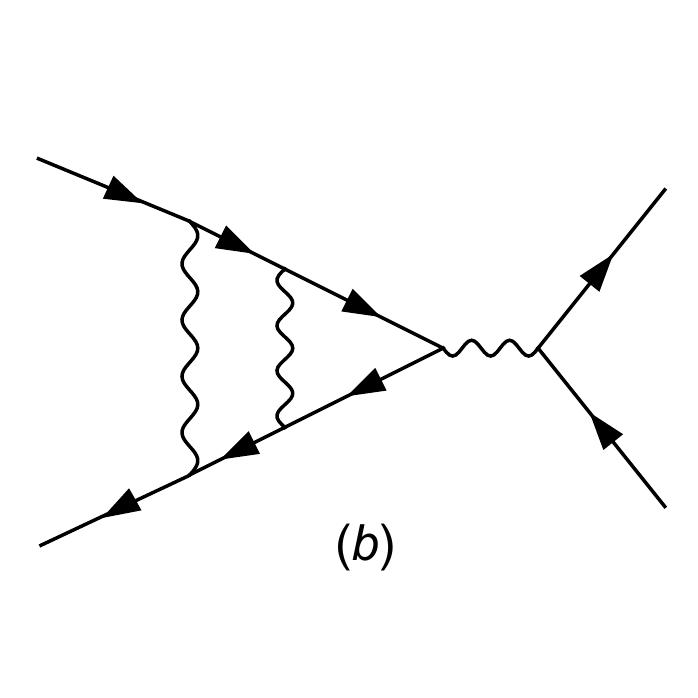}
\end{minipage}
\begin{minipage}{0.33\textwidth}
\centering
\includegraphics[width=0.9\textwidth]{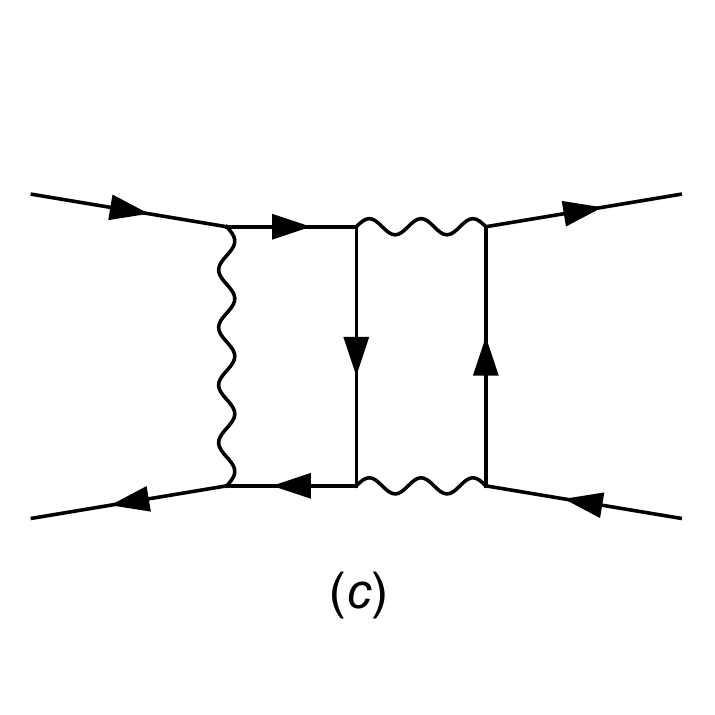}
\end{minipage}
\begin{minipage}{0.32\textwidth}
\centering
\includegraphics[width=0.95\textwidth]{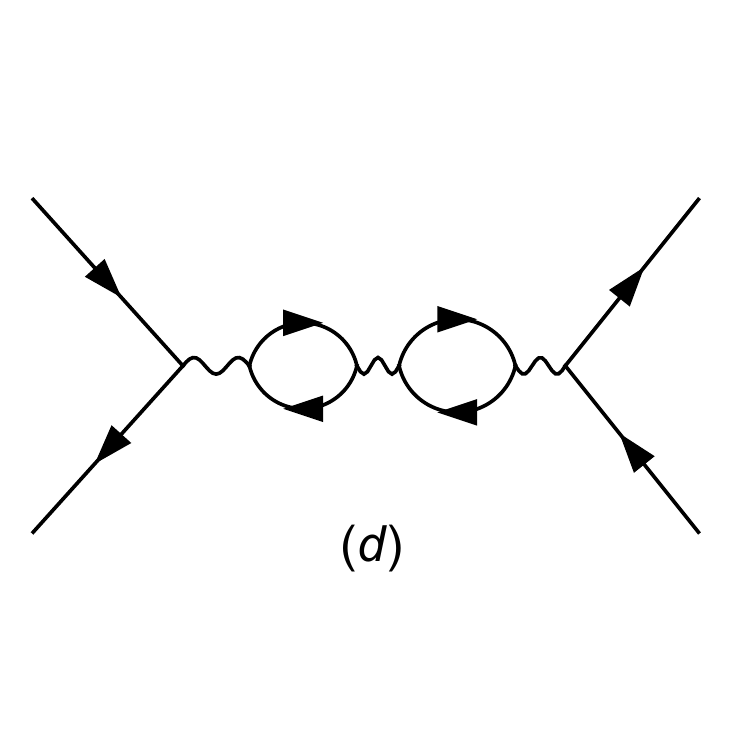}
\end{minipage}
\begin{minipage}{0.32\textwidth}
\centering
\includegraphics[width=0.95\textwidth]{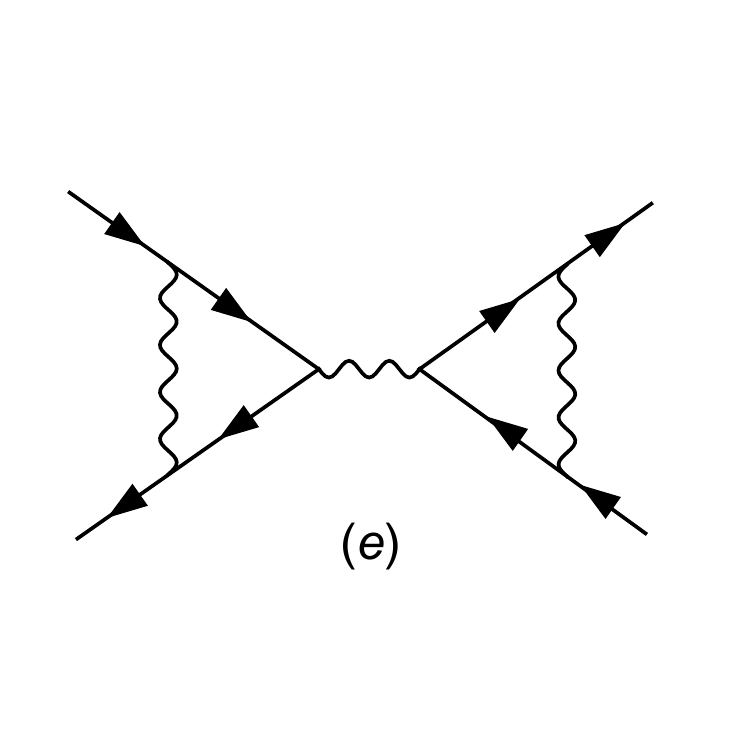}
\end{minipage}
\begin{minipage}{0.32\textwidth}
\centering
\includegraphics[width=0.95\textwidth]{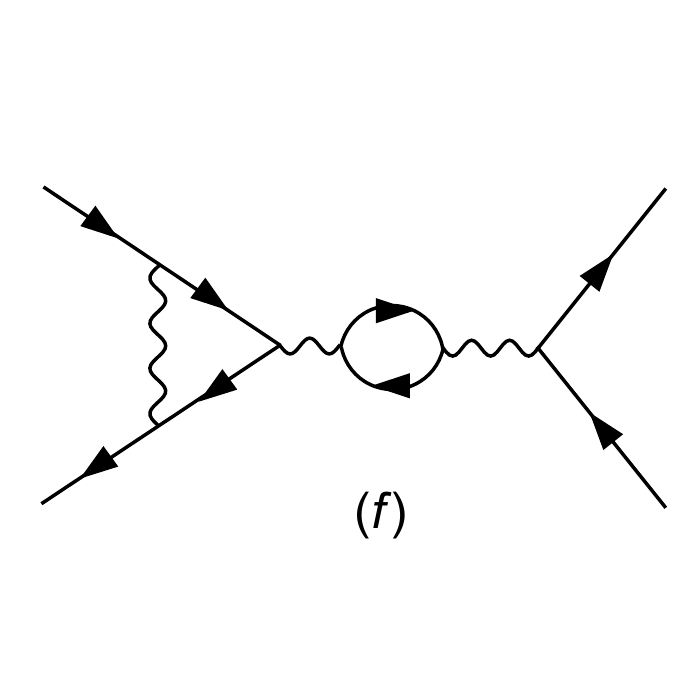}
\end{minipage}
\caption{\label{fig:sample2Ldiagrams} Sample two-loop Feynman diagrams: 1-particle irreducible and factorisable one-loop $\times$one-loop contributions. }
\end{figure}

In Figure \ref{fig:sample2Ldiagrams} are presented some examples of two-loop Feynman diagrams contributing to the bare interference term $\langle \unM^{(0,0)} | \unM^{(0,2)} \rangle$. $(a),\,(b),\,(c)$ are sample diagrams for each 1-particle irreducible (1PI) category in which we can group the diagrams, respectively self-energies, vertices and boxes 1PI diagrams. In contrast, $(d),\,(e),\,(f)$ represent sample diagrams for the factorisable contributions given by the iteration of 2 one-loop diagrammatic corrections. 
The cancellation of the UV divergences stemming from the bare diagrams follows according to a standard renormalisation procedure. The resulting amplitudes are expressed in terms of renormalised parameters, in turn related to measurable quantities.
We write the UV-renormalised amplitude as a Laurent expansion in $\epsilon$, and we keep only terms up to 
$\mathcal{O}(\epsilon^2)$ or $\mathcal{O}(\epsilon^0)$, at one- or two-loop respectively, in order to include all the second-order finite contributions when we perform the IR subtraction described in Section~\ref{sec:infra}.

\subsection{Ultraviolet renormalisation}
\label{sec:UV}
The SM renormalisation up to second order in the electroweak interaction has been discussed in \cite{Denner:1994xt,Denner:2019vbn,Dittmaier:2021loa,Actis:2006ra,Actis:2006rb,Actis:2006rc} both for the on-shell and the $\overline{MS}$ renormalisation scheme. The electric charge renormalisation at \oaa has been presented in detail in \cite{Degrassi:2003rw} while in \cite{Bonciani:2021okt} it is possible to find the analytical expressions of the fermion mass and wave-function counterterms in QED for the case of one internal mass. Our renormalisation procedure within \ocean~ is realized with a dedicated semi-analytical evaluation of the coupling and fermions renormalisation constants, whose definitions follow from the 2-point Green functions of the theory and can be found in Appendix \ref{app:renrel}. The generalisation of this computational approach to the complete two-loop EW case is straightforward.

We write the virtual amplitude for the production of a pair of massive electrons as
\begin{equation}
\label{eq:expansion}
    |\unM (u\bar u\to e^+e^-)\rangle = 4\pi\hat\alpha \sum_{j=0}^{\infty} \hat\aem^j |\unM^{(0,j)}(s,t,\hat{m}_{e},\hat{m}_{f})\rangle\;,
\end{equation}
where $|\hat{\cal M}^{(0,j)}\rangle$ is the bare amplitude at $j^{th}$ perturbative order in $\hat{\alpha}$, and all the parameters (the electric charge $\hat e$, the electron mass $\hat m_e$ and the mass $\hat m_f$ of a generic fermion $f$) and fields are treated as bare quantities, labeled with a hat. 
The renormalised amplitude is obtained by the replacement of all the bare quantities with their renormalised counterparts:
$\hat{e}\to e+\delta e$,  
$\,\,\hat{m}_{e}\to m_{e}+\delta m_{e}$, $\hat{m}_{f}\to m_{f}+\delta m_{f}$, and the rescaling $\hat{\psi}\to Z_\psi^{\frac12}\psi=(1+\delta Z_\psi)^{\frac12}\psi$ for each external fermionic field, where we consider $\hat\alpha=\hat e^2/(4\pi)$ and $\alpha=e^2/(4\pi)$, 
\begin{align}
|{\cal M}(u\bar u\to e^+e^-)\rangle~&=~
Z^{\frac12}_u Z^{\frac12}_{\bar u} Z^{\frac12}_{e^+} Z^{\frac12}_{e^-}~16\pi^2~\times\nonumber\\
&~~~~~\sum_{j=0}^\infty \left(\frac{e+\delta e}{4\pi}\right)^{2j+2} |\hat{\mathcal{M}}^{(0,j)}(s,t,m_{e}+\delta m_e,m_f+\delta m_f)\rangle\nonumber\\
&= \ 4\pi\alpha\sum_{j=0}^\infty \left(\frac{\alpha}{4\pi}\right)^{j} |\mathcal{M}^{(0,j)}(s,t,m_{e},m_f)\rangle\;.
\label{eq:amprenormalised}
\end{align}
Since not only the bare amplitude, but also the counterterms are computed in perturbation theory, we expand the renormalised amplitude and  collect all the terms of the same order in the coupling constant, checking the cancellation of the UV divergences.
The result can be eventually cast as an expansion in the renormalised coupling $\alpha$, where $|\mathcal{M}^{(0,j)}\rangle$ includes diagrammatic and counterterm contributions and is UV finite.

\begin{figure}
\begin{minipage}{0.32\textwidth}
\centering
\includegraphics[width=0.95\textwidth]{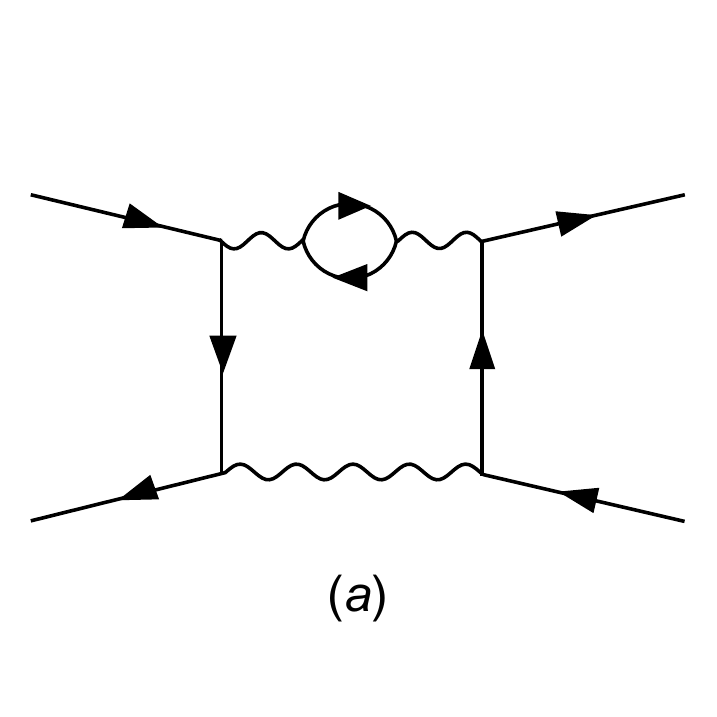}
\end{minipage}
\begin{minipage}{0.32\textwidth}
\centering
\includegraphics[width=0.95\textwidth]{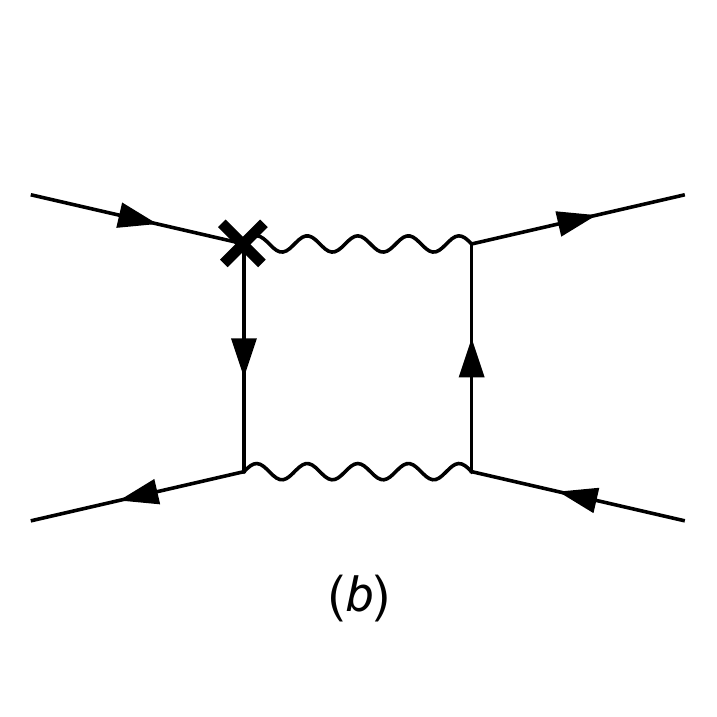}
\end{minipage}
\begin{minipage}{0.32\textwidth}
\centering
\includegraphics[width=0.95\textwidth]{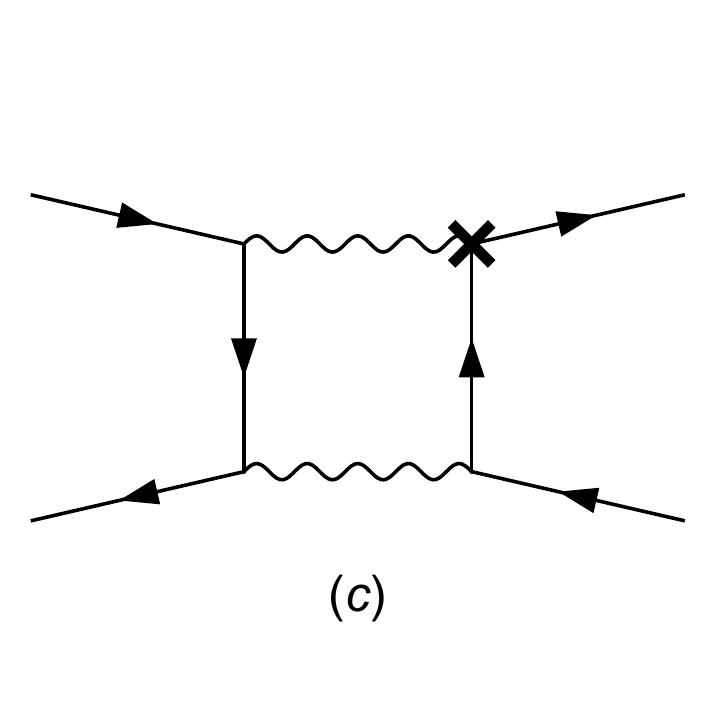}
\end{minipage}
\caption{\label{fig:fermionloopandCTs} Sample Feynman diagrams with the insertion of a closed fermionic loop along a photon line and the associated electric charge counterterm diagrams.}
\end{figure}

At first order, we identify three subsets of Feynman diagrams, the corrections to the tree-level photon propagator, the initial (final) vertex corrections and the associated wave function (WF) factors on the external quark (electron) lines, the box diagrams. 
Each of these three groups of diagrams is UV finite, after the inclusion of the one-loop counterterms: 
the boxes are UV finite by power counting, 
the photon vacuum polarization is by construction made finite by the electric charge counterterm;
the Ward identities guarantee the UV finiteness of the combination of the vertex corrections with the WF constants of the corresponding external lines.
Additional poles in the dimensional regularization parameter $\epsilon$ are however still present, because of the IR divergent behaviour of the vertex and box corrections.

At the second perturbative order, the finiteness of the self-energy corrections to the tree-level photon propagator follows from the definition of the electric charge counterterm, once all the factorisable one-loop contributions have been properly included. 
Like in the one-loop case, the UV-finiteness of the sum of vertex and WF contributions can not be directly checked, because of the presence of IR poles, also expressed with the regulator $\epsilon$.
We identify two distinct groups of second-order 1PI diagrams: those with the exchange of two internal photons (like for instance diagrams $(b)$ and $(c)$ in Figure \ref{fig:sample2Ldiagrams}), and those with the insertion, along an internal photon line of a one-loop diagram, of a closed fermionic loop (like for instance diagram $(a)$ in Figure \ref{fig:fermionloopandCTs}).
The latter features an UV divergence which is subtracted by the one-loop electric charge counterterm applied to the corresponding one-loop amplitude (diagrams $(b)$ and $(c)$ in Figure \ref{fig:fermionloopandCTs}).

Since the IR subtraction terms discussed in Section \ref{sec:infra} have been derived in the $\overline{MS}$ renormalisation scheme for the electric charge, we have to adopt it also in the evaluation of the virtual corrections to ensure a consistent cancellation of the IR divergencies.
While our final result will thus be expressed in terms of the coupling coupling $\alpha_{\overline{MS}}(\mu_R)$, evaluated at the renormalisation scale $\mu_R$, it can also be expressed in terms of the on-shell fine structure constant $\alpha(0)$, with a finite renormalisation at the end of the calculation, as it is discussed in detail in Section~\ref{sec:schemes}.
The fermion masses and the fermion wave-function counterterms are instead always computed in the on-shell scheme.

We remark that a change in the electric charge renormalisation scheme has a non trivial impact on the pattern of the subtraction of the IR poles. If we consider as an example the two diagrams $(b)$ and $(c)$ in Figure \ref{fig:fermionloopandCTs}, we can easily recognize the fact that a finite additional contribution to the counterterm insertion is a constant which multiplies the entire underlying one-loop diagram. The latter features soft and collinear IR poles, which are multiplied by this additional factor. The cancellation of these extra poles is possible only with a corresponding contribution from the real emission diagrams, or equivalently from the subtraction term.

\section{Results}
\label{sec:results}

The calculation presented in this paper illustrates with a non trivial example the application of \ocean, a new complete workflow for the evaluation of two-loop EW corrections. It provides in particular the validation of a new semi-analytical approach to handle the cancellation of UV and IR divergences, represented by poles in dimensional regularization.

\subsection{Automatic generation and evaluation of the two-loop virtual corrections}

The computational framework \ocean~ is a collection of codes and packages that provides, to an experienced user, all the necessary tools to face the challenges of the calculation of a multi-loop electroweak amplitude in a consistent way.
The main steps of these computations are:
the calculation of the squared matrix element which contains all the interference terms between the Feynman diagrams that contribute to the process; the evaluation of the Feynman loop integrals present in the amplitude; and, finally, the cancellation of the UV and IR divergences via the renormalisation and subtraction procedures.

The evaluation of the squared matrix elements is implemented, within \ocean, in the Mathematica package \abiss~ (Amplitudes Builder, Interference Solver and Simplifier).
\abiss~ 
expresses the final result as a combination of MIs, each multiplied by a rational function of the kinematical invariants and masses of the process,
with a workflow which can be described schematically by the following points:
\begin{enumerate}
    \item The generation of the Feynman diagrams of the chosen scattering process is performed with {\tt FeynArts} \cite{Hahn:2000kx}, which supports several model files for the SM and its extensions.
    \item 
    The Feynman diagram with the insertion of UV counterterms are generated with a custom model file for {\tt FeynArts} which provides the counterterms in symbolical form. Their numerical values, as a Laurent series in the dimensional regularization parameter $\epsilon$, can be computed together with all the other diagrams, or they can be read from any external counterterm library.
    \item The evaluation of the interference between an amplitude and its hermitian conjugate\footnote{We can in full generality consider an amplitude and a projector.}, including the sum over all the external polarizations, is performed by dedicated routines in \abiss. The result is a list of expressions, Lorentz scalars, including Feynman loop integrals and coefficients, functions of kinematical invariants and masses.
    \item All the Feynman integrals are classified in terms of integral families. 
    The user can propose a preferred choice of integral families, in terms of which \abiss~ tries to write all the Feynman integrals present in the computation\footnote{Different routings of the loop momenta and trivial symmetry relations are tested, to match the momenta with those of the definition of the integral family.}. If the assignment of a given Feynman integral to one of the available families is not possible, \abiss~ automatically completes the tentative set,
    using the information of the new Feynman integral to introduce the new family. 
    At the end of this classification step, the initial  information relevant for the reduction to a minimal set of so called MIs is available and automatically organized in a group of dedicated directories.
    \item The output of the previous point is ready for the external run of any code able to generate all the relations useful to express a generic element of an integral family as a linear combination of MIs. Such relations include IBP and Lorentz identities. Within \ocean, we consider as reference code for this task \kira\, \cite{Maierhofer:2017gsa,Klappert:2020nbg,Klappert:2019emp,Lange:2025fba}.
    The outcome of the run can be imported back to the directories already set-up by \abiss.
    \item With the help of the reduction rules computed in the previous point, \abiss~ is able to organize the squared matrix element as a sum over the $N_{MI}$  MIs ${\cal I}_j$, with appropriate coefficients $c_j$. For example, in the case of the two-loop contributions,
    \begin{equation}
    \langle {\cal M}^{(0)} | {\cal M}^{(0,2)} \rangle \,
    =
    \sum_{j=1}^{N_{MI}} c_j\, {\cal I}_j\,\, .
    \label{eq:sumcoeffMIs}
    \end{equation}
    Both coefficients and integrals depend on the kinematical invariants and on the masses of the particles of the model. 
\end{enumerate}

The evaluation of the MIs which have been identified in the first stage is handled within \ocean~ with the public Mathematica package \seasyde~\cite{Armadillo:2022ugh,Armadillo:2025mvu}, following a simple workflow:
\begin{enumerate}
    \item For each integral family, we compute the systems of differential equations satisfied by its MIs, with respect to the kinematical invariants. For this task, several tools are publicly available, like e.g. {\tt LiteRed} \cite{Lee:2013mka} or {\tt REDUZE} \cite{vonManteuffel:2012np}.
    \item We compute the boundary condition values for each MI with the code \amflow\, \cite{Liu:2022chg}.
    \item We solve the system of differential equations with \seasyde.
\end{enumerate}
\seasyde\, 
solves the system of differential equations with the
 series expansion procedure \cite{Moriello:2019yhu,Hidding:2020ytt}. The solution is studied in the complex plane, with an original algorithm for the analytical continuation of the solutions. The efficient evaluation of a series of points in one single run is relevant for the preparation of numerical grids which sample the whole phase space of the process. The numerical results have arbitrary precision, which can be selected by the user and is achieved with an appropriate number of terms in the power series used to represent the solution.

The last steps of the calculation are handled by internal routines within \ocean:
the combination of the diagrammatic and counterterm contributions, 
to achieve the UV finiteness of the amplitude; 
the subtraction of the IR poles from the UV renormalised interference terms, according to the strategy described in Section \ref{sec:infra}. The sequences of points, covering the whole phase space, computed by Seasyde for the MIs and within \abiss~for their rational coefficients, express the finite contribution of the virtual corrections, relevant for phenomenological studies. One possible strategy is the preparation of grids which allow the evaluation of the correction at any phase-space point via interpolation, with excellent accuracy, in negligible time. This approach has been applied, for example, in the context of the mixed QCDxEW corrections in~\cite{Armadillo:2022bgm,Armadillo:2024nwk}.

The sequence of computational steps described above, taking place within the \ocean\, framework, is fully general and applies to the evaluation of the squared matrix element of an arbitrary scattering process at a given perturbative order.

\subsection{Explicit QED results and control over the cancellation of the IR poles} \label{subsec:Explicit QED results}

In the following, we present our results for the squared two-loop QED virtual amplitudes $\langle {\cal M}^{(0)} | {\cal M}^{(0,2)}\rangle$ for the process $u\bar u\to e^+e^-$.
We identify in total 11 families relevant for the description of two-loop box topologies\footnote{We count only once the insertion of a closed fermionic loop, treating the fermion mass as one generic parameter.}.
In the classification of the vertex corrections we identify 6 new families, 
while in the analysis of the self-energy diagrams we find 2 new families.

The fully diagrammatic approach allows us to perform intermediate checks of the calculation. 
We find a finite result for the self-energy corrections to the tree-level photon propagator, after the introduction of the one- and two-loop electric charge counterterms.
We check the cancellation of one pole in $\epsilon$, when we combine a two-loop diagram with one fermion loop insertion together with the charge renormalisation of the adjacent vertices in the corresponding one-loop diagram.

The UV-renormalised results described in the previous Section still feature IR divergences, which appear
as poles in the dimensional regularization parameter $\epsilon$, but also as logarithms of the lepton mass in the case of the final state collinear divergences. 
It is a Laurent series in the dimensional regularization parameter, 
and the coefficients of the various powers of $\epsilon$ are numbers with arbitrary precision.
In Table \ref{tab:ircanc} we present the value of the coefficients of the different powers of $\epsilon$, 
for the diagrammatic part and for the the two contributions to the subtraction operator, factorising the one-loop and Born squared amplitudes respectively. These results are obtained using the following values for the masses and Mandelstam invariants:
\begin{center}
 \begin{tabular}{l l} 
 \hline\hline
  $s=10000~\mathrm{GeV}^2$ &\hspace{20pt} $m_e$ = 0.510998 MeV\\
  $t=-2500~\mathrm{GeV}^2$ &\hspace{20pt} $m_\mu$ = 0.10566 GeV\\
  $\mu_R=m_Z$=91.1535 GeV  &\hspace{20pt} $m_\tau$ = 1.77686 GeV\\
   &\hspace{20pt} $m_t$ = 172.69 GeV\\
   \hline\hline
 \end{tabular} 
\end{center}
\begin{table}[t]
    \centering
    \begin{tabular}{c|S[table-format = 7.33]}
    \hline
        \multirow{3}{*}{$\epsilon^{-4}$} 
         &0.43895747601285225040892043215096  \\
         &-0.43895747601285225040892043215096 \\
         &0 \\
    \hline
        \multirow{3}{*}{$\epsilon^{-3}$} 
        &-50.507936715422813794673578353212  \\
        &50.507936715422813794673578353212 \\
        &0 \\
    \hline
        \multirow{3}{*}{$\epsilon^{-2}$} 
         &729.67339746440682380540184060883  \\
         &-1334.17941387822635237175428142592 \\
         &604.50601641381952856635244081709 \\
    \hline
        \multirow{3}{*}{$\epsilon^{-1}$} 
        &26528.66234466048956681205858967355  \\
        &-463.09023418725937944723920907369 \\
        &-26065.57211047323018736481938059986 \\
    \hline
        \multirow{3}{*}{$\epsilon^{0}$} 
        &401847.4473451308127296540393826352  \\
        &4376.8127679941628242540225839402 \\
        &-308980.3885355040551980137489252684 \\
    \hline 
        {$2\text{Re}\langle {\cal M}^{(0,0)}|{\cal M}^{(0,2),fin}\rangle$}
        &97243.8715776209203558943130413070 \\
    \hline
    \end{tabular}
    \caption{Numerical coefficient, computed with at least 32 significant digits, of the IR poles present in the different contributions to Equation~(\ref{eq:subtracted}), for $s=10000 \ \text{GeV}^2$ and $t=-2500 \ \text{GeV}^2$. For each pole, the first row refers to $2\text{Re}\langle {\cal M}^{(0,0)}|{\cal M}^{(0,2)}\rangle$, the second one to $-{\cal I}^{(0,2)}\; 2\text{Re} \langle {\cal M}^{(0,0)}|{\cal M}^{(0,0)}\rangle$ and the third one to $-{\cal I}^{(0,1)} \; 2\text{Re}\langle {\cal M}^{(0,0)}|{\cal M}^{(0,1),fin}\rangle$. The last row is the final result of $2\text{Re}\langle {\cal M}^{(0,0)}|{\cal M}^{(0,2),fin}\rangle$.}
    \label{tab:ircanc}
\end{table}
In the evaluation of the individual contributions to the amplitude, we encounter some terms whose value is several orders of magnitude larger than 1; we can inspect them all in absolute value and define $Q=\log_{10}(x)$, with $x$ the largest term of the calculation.
We consider verified the cancellation of the IR poles if, for any given number $P$ of significant digits requested in the arbitrary precision evaluation framework, with $P>Q$, the sum of the coefficients yields $10^{-P+Q}$, which is our explicit representation of zero. The results in Table \ref{tab:ircanc} have been obtained with a precision request $P=50$, while we observe a precision loss due to internal cancellations by $Q=15$ orders of magnitude, eventually achieving control over 35 digits. 
The IR cancellation can be verified in greater detail by isolating subsets of contributions proportional to the same power of the electric charges of the initial state quark, final state lepton, and of the fermion running in the internal loops; each subset is separately IR finite.

Whenever possible, the precision losses can be reduced with a convenient choice of the MIs basis. 
In this computation we use the basis of Master Integrals automatically generated by the combined application of {\tt Feynarts}, {\tt ABISS} and \kira,
in order to test the efficiency of our code also with a suboptimal choice of Master Integrals. 
For more challenging calculations, as the full EW case, it is possible to improve such choice with the help of packages like for instance {\tt Canonica} \cite{Meyer:2017joq} or {\tt LIBRA} \cite{Lee:2020zfb}.
The numerical control over the pole cancellation is the same as the one obtained using  closed form expressions in terms of special functions for the Feynman integrals, given the possibility to push the precision to an arbitrary level.
The solution of the MIs with a series expansion approach offers on the other hand a significant advantage for all those cases where a closed form solution is not viable: IR-finite integrals with several internal massive lines, but also IR-divergent integrals with 2 or 3 internal massive lines, which require a non trivial dedicated effort to be expressed in closed form \cite{Heller:2019gkq}.

\begin{figure}[t]
    \centering
    \includegraphics[width=0.75\linewidth]{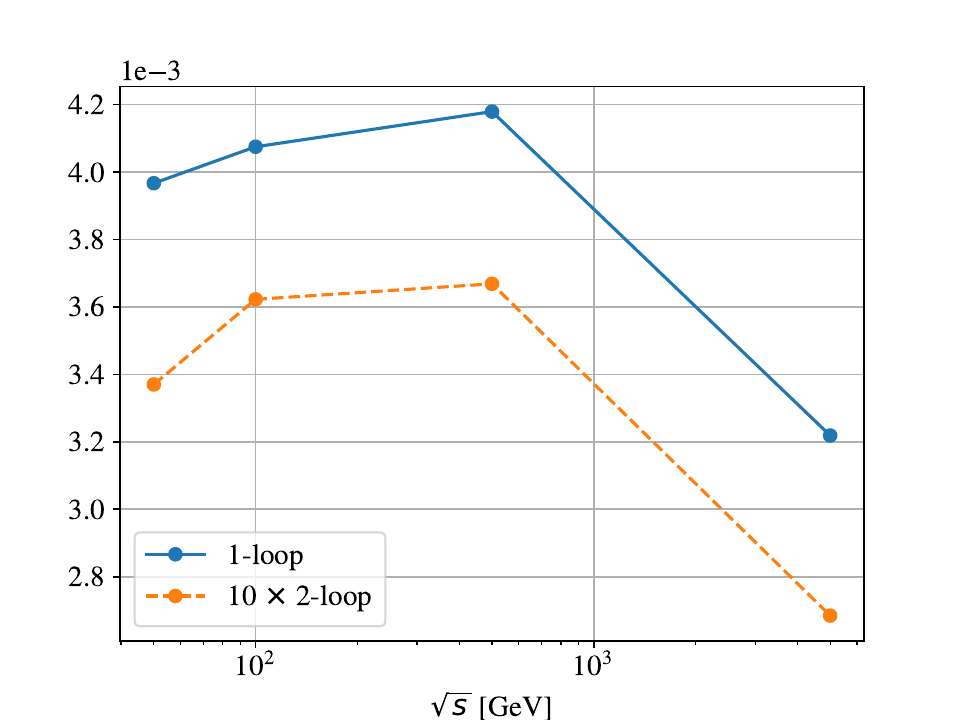}
    \caption{We plot the contribution to $\langle\mathcal{M}|\mathcal{M}\rangle$ coming from one- and two-loop diagrams for different values of $\sqrt{s}$ and $t=-s/4$.
    In particular, in solid blue we present $(4\pi\alpha_{\overline{\text{MS}}}^3)\;2\text{Re}\langle {\cal M}^{(0,0)}|{\cal M}^{(0,1),fin}\rangle$, while in dashed orange $\alpha_{\overline{\text{MS}}}^4\; 2\text{Re}\langle {\cal M}^{(0,0)}|{\cal M}^{(0,2),fin}\rangle$, with $\alpha_{\overline{\text{MS}}}=1/128$. The two-loop contributions have been multiplied by a factor $10$ to improve the readability.}
    \label{fig:M2fin}
\end{figure}

In Table \ref{tab:ircanc} we present also the result for the finite part of the interference term, which is in perfect agreement with the values presented in \cite{Bonciani:2021okt}.
In Figure \ref{fig:M2fin}, we present additional values for the finite part of the interference term, calculated for different values of $\sqrt{s}$, whose behaviour displays the non trivial interplay of the different powers of Sudakov logarithms present in the virtual corrections.

\subsection{Impact of a change of scheme} \label{sec:schemes}
The results discussed in Section \ref{sec:process} are computed in terms of the $\overline{MS}$-renormalised electric charge  $\alpha_{\overline{MS}}(\mu_R)$. 
The UV renormalised and IR subtracted squared amplitude, including the corrections up to second order, reads:
\begin{align}\label{eq:amplMSbar}
    \langle \M |\M \rangle =  \ & 16\pi^2 \alpha_{\overline{MS}}^2(\mu_R) \langle \M^{(0,0)}|\M^{(0,0)}\rangle 
    \\\nonumber 
    &+ 4\pi \alpha_{\overline{MS}}^3(\mu_R) \left[2 \text{Re} \left( \langle \M^{(0,0)}|\M^{(0,1),fin}\rangle\right) \right]  
    \\ \nonumber
    &+\alpha_{\overline{MS}}^4(\mu_R) \left[ 2\text{Re} \left( \langle \M^{(0,0)}|\M^{(0,2),fin}\rangle\right) + \langle \M^{(0,1),fin}|\M^{(0,1),fin}\rangle \right] ~+~
    \cdots
    \\\nonumber
    = \ & \alpha_{\overline{MS}}^2(\mu_R) \ \bar{\mathcal{A}}_{00} + \alpha_{\overline{MS}}^3(\mu_R) \ \bar{\mathcal{A}}_{01} + \alpha_{\overline{MS}}^4(\mu_R) \ \left[ \bar{\mathcal{A}}_{02} + \bar{\mathcal{A}}_{11}\right]~+~
    \cdots\;, 
\end{align}
where the amplitudes $| {\cal M}^{(0,1),fin}\rangle$, $| {\cal M}^{(0,2),fin}\rangle$ and the abbreviations $\bar{\cal A}_{ij}$ depend on the renormalisation scale $\mu_R$.
Another choice to express the renormalised coupling is given by the on-shell scheme definition,   at zero momentum transfer, $\alpha(0)$, from the study of the Thomson scattering. This second alternative provides an accurate description of the real-photon emission spectra, and it is thus of high phenomenological relevance.
It is interesting to re-express Equation~(\ref{eq:amplMSbar}) in terms of the fine structure constant $\alpha(0)$, via a finite renormalisation. 
The relation between the on-shell and $\overline{MS}$ renormalised electric charges is given by \cite{Degrassi:2003rw}:
\begin{equation}
    \alpha_{\overline{MS}}(\mu_R)
    ~=~
    \frac{\alpha(0)}{1-\Delta \alpha_{\overline{MS}}(\mu_R^2)} \; .
\label{eq:onshellMSbar}
\end{equation}
The definitions of the coefficients $\Delta\alpha^{(i)}_{\overline{MS}}(\mu_R^2)$ are reported in Appendix \ref{app:vp}, 
with the first order containing the hadronic part $\Delta\alpha_{\text{had}}(\mu_R^2)$, necessary to express the light-quarks contribution. 
This quantity is extracted from the experimental data of the cross section $e^+ e^- \rightarrow \text{hadrons}$ and, in the case of $\mu_R=m_Z$, it is possible to use the determination from \cite{Proceedings:2019vxr}:
$\Delta\alpha_{\text{had}}(m_Z^2) = 0.027572 \pm 0.000359$.
In the following we perform a technical comparison to test the perturbative convergence of the results when increasing the order of the perturbative approximation. 
Expanding Equation~(\ref{eq:onshellMSbar}) one gets:
\begin{equation}
    \alpha_{\overline{MS}}(m_Z)
    = \
    \alpha(0) \left[ 1+\alpha(0) \Delta \alpha^{(1)}_{\overline{MS}}(m_Z^2) + \alpha(0)^2 \left( \left( \Delta \alpha^{(1)}_{\overline{MS}}(m_Z^2) \right)^2  + \Delta \alpha^{(2)}_{\overline{MS}}(m_Z^2) \right) \right]\; .
\label{eq:onshellMSbar_expanded}
\end{equation}
Replacing the couplings in Equation~(\ref{eq:amplMSbar}) with Equation~(\ref{eq:onshellMSbar_expanded}), we obtain:
\begin{align}    
    \langle \M | \M \rangle =&
   \  \alpha(0)^2\,\bar{\mathcal{A}}_{00}~+~
     \alpha(0)^3\,
    \left[\bar{\mathcal{A}}_{01} + 2 \Delta\alpha^{(1)}_{\overline{MS}}(m_Z^2)\, \bar{\mathcal{A}}_{00}\right]
    \label{eq:xseconshell}\\\nonumber
    & + \alpha(0)^4\,
    \Big[
    \bar{\mathcal{A}}_{02} + \bar{\mathcal{A}}_{11} + 
    3 \Delta\alpha^{(1)}_{\overline{MS}}(m_Z^2)\, \bar{\mathcal{A}}_{01}  
    \\\nonumber&
    + 
    2 \Delta\alpha^{(2)}_{\overline{MS}}(m_Z^2)\, \bar{\mathcal{A}}_{00} +
    3 \left(\Delta\alpha^{(1)}_{\overline{MS}}(m_Z^2)\right)^2\, \bar{\mathcal{A}}_{00}
    \Big]~+~
    \cdots\nonumber
     \end{align}
\begin{table}
    \centering
    \begin{tabular}{c|c|c|c}
         Highest order & $\alpha(0)$ scheme & $\alpha_{\overline{MS}}(m_Z)$ scheme & $\Delta~(\%)$ \\[3pt]
        \hline & & &\\[-10pt]
        tree-level &   46.742   &  53.546 &  12.707 \\[3pt]
        one-loop &   86.610  &  94.297 &  8.152 \\[3pt]
        two-loop &   97.158   &  97.919 &  0.777 \\[3pt]
        \hline
    \end{tabular}
    \caption{Numerical results of the squared virtual amplitude at $\mathcal{O}\left( \alpha^2\right)$, $\mathcal{O}\left( \alpha^3\right)$ and $\mathcal{O}\left( \alpha^4\right)$ in the two   schemes for the renormalised electric charge, with $\alpha_{\overline{\text{MS}}}(m_Z)=1/128$ and $\alpha(0)=1/137$, in units of  $10^{-4}$. The last column shows the percentage difference $\Delta= 100\,\left|\langle \M | \M \rangle_0~/~\langle \M | \M \rangle_{\overline{MS}} -1\right|$.}
    \label{tab:inputschemes}
\end{table}
In Table \ref{tab:inputschemes} we compare the sum of the contributions to the squared matrix element, using as renormalised electric charge $\alpha_{\overline{MS}}(m_Z)$ or $\alpha(0)$\footnote{Note that in order to use the formulae in~\cite{Degrassi:2003rw}, we have to account for the different normalisation of the Feynman integrals by a factor $S_\epsilon$.}.
We consider different approximations: tree-level only,  tree-level and one-loop corrections, the sum of tree-level, one-loop and of the two-loop $\langle \M^{(0,0)}|\M^{(0,2),fin}\rangle$ term.
At tree level the difference between the results is purely due to the ratio of the coupling constants 
$\left(\alpha_{\overline{MS}}(m_Z)/\alpha(0)\right)^2$.
In the one-loop virtual corrections there are two subsets: the self-energy corrections together with the electric charge counterterms
and the group composed by vertex and box corrections. The former, combined with the tree level, features in the two schemes much closer results compared to the tree-level case. The latter on the other hand are still evaluated with tree level couplings and, given the large difference between the couplings in the two schemes, they yield a difference between the results at the 8\% level. 
It is thus interesting to observe the impact of the two-loop corrections, which include the renormalisation of the couplings present in the one-loop diagrams. Thanks to this additional step, also the one-loop vertices and boxes are evaluated with effective couplings closer in value, reducing the distance between the two schemes at sub percent level.

\section{Conclusions}
\label{sec:conclusions}
We have presented in this paper the evaluation in QED of the two-loop virtual corrections to the process of  production of a massive lepton-pair, in quark-antiquark annihilation.
These results illustrate a general systematic approach to the evaluation of the complete two-loop EW corrections to the scattering processes relevant for the precision physics program at the LHC. The tools needed to accomplish this ambitious task are coherently organized in the \ocean\, framework. One of the major technical issues, the evaluation of the Feynman integrals, is solved with a solution by series expansions, which provides results with arbitrary precision and a complete control over the cancellation of the UV and IR divergences.
The validation of these QED results with the literature supports the robustness of the approach, in a stress case from the point of view of the numerical cancellations and of potential precision losses.

\section*{Acknowledgments}
We would like to thank Roberto Bonciani,
for several interesting discussions. We would like to thank Pierpaolo Mastrolia, Manoj Mandal, William Torres Bobadilla and Jonathan Ronca for providing us with the numerical values of the amplitude from Ref.~\cite{Bonciani:2021okt}.
T.A. is a Research Fellow of the Fonds de la Recherche Scientifique – FNRS.
The work of S.D. received support from the European Union (ERC, MultiScaleAmp, Grant Agreement No. 101078449) and the FWO (contract No. 1227426N). Views and opinions expressed are however those of the author(s) only and do not necessarily reflect those of the European Union or the European Research Council Executive Agency. Neither the European Union nor the granting authority can be held responsible for them.
A.V. was partially supported by the Project W2441004 of the National Natural Science Foundation of China.
\newpage
\appendix
\section{Renormalisation relations}
\label{app:renrel}

We report here the basic definitions that we have followed for the semi-analytical computation of the counterterms, needed to obtain the complete two-loop renormalised amplitude.

\subsection{Electric charge renormalisation within the \texorpdfstring{$\overline{MS}$}{MS} scheme}

The renormalisation of the electric charge has been carried out following the same scheme for both the diagrammatic contributions and the subtraction operators. The relation between the bare electric charge and the renormalised one is defined up to the second order in $\alpha$ by introducing the counterterms
\begin{equation}
    e_0 = \ e+\delta e 
    = \ e \ Z_e= \ e \left( 1+\delta Z_e^{(1)} + \delta Z_e^{(2)} + \mathcal{O}(\alpha^3)\right)\;.
\end{equation}
Following the on-shell renormalisation scheme, the Ward Identity associated to the residual $U(1)_{em}$ gauge invariance  ensures that the renormalisation of the electric charge follows entirely from the transverse photon-photon self-energy $\Sigma_{T}^{AA}(q^2) $\cite{Denner:1994xt,Degrassi:2003rw,Dittmaier:2021loa}, via the renormalisation condition on the $\gamma f f$ vertex that relates $Z_e$ with the photon wavefunction counterterm $Z_{AA}=1+\delta Z_{AA}^{OS}=1+\delta Z_{AA}^{OS(1)}+\delta Z_{AA}^{OS(2)}+\cdots$:
\begin{equation} \label{dZedZAA relation}
    Z_e^{OS} = \ \frac{1}{\sqrt{Z_{AA}^{OS}}} = \ \frac{1}{\sqrt{1+\delta Z_{AA}^{OS(1)}+\delta Z_{AA}^{OS(2)}+\dots}} \; ,
\end{equation}
with
\begin{equation} \label{renorm condition dZAA}
    \delta Z_{AA}^{OS} = - \dfrac{\partial\Sigma_{T}^{AA}(q^{2})}{\partial q^{2}} \bigg|_{q^{2}=0} \; .
\end{equation}
In the $\overline{MS}$ scheme, only the divergent terms of the self-energies are kept in the definitions of the counterterms, and an additional dependence on the renormalisation scale $\mu_R$ and on the regularisation 't Hooft scale $\mu$ is introduced together with an overall factor $S_\epsilon=(4\pi e^{-\gamma_E})^\epsilon$:
\begin{align}
\delta Z_e^{\overline{MS}(1)}&= \ S_\epsilon \left( \frac{\mu^{2}}{\mu_R^{2}}\right)^\epsilon\left[-\frac{1}{2} \delta Z_{AA}^{OS(1)}\big|_\text{poles}\right] \; , \\ \nonumber
\delta Z_e^{\overline{MS}(2)}&= \ S_\epsilon^2 \left( \frac{\mu^{2}}{\mu_R^{2}}\right)^{2\epsilon} \left[-\frac{1}{2} \delta Z_{AA}^{OS(2)}\big|_\text{poles}+\frac{3}{8} \left(\delta Z_{AA}^{OS(1)}\big|_\text{poles} \right)^2\right]  \; .
\end{align}
We have checked our results to be in agreement with an arbitrary number of digits with the expressions provided in \cite{Bonciani:2021okt}, that we that we report here for completeness after switching to our notation:
\begin{align}
    \delta Z_e^{\overline{MS}(1)} &= \ \frac{\alpha(\mu_R)}{4\pi} S_\epsilon\left( \frac{\mu^{2}}{\mu_R^{2}}\right)^\epsilon\left[ \frac{2}{3\epsilon} \sum_{k} N_k Q_k^2 \right] \; , \\ \nonumber
   \delta Z_e^{\overline{MS}(2)} &= \ \left(\frac{\alpha(\mu_R)}{4\pi}\right)^2 S_\epsilon^2\left( \frac{\mu^{2}}{\mu_R^{2}}\right)^{2\epsilon}\left[ \frac{2}{3 \epsilon^2} \left(\sum_{k} N_k Q_k^2\right)^2 + \frac{1}{\epsilon} \sum_{k} N_k Q_k^4\right] \; .
\end{align}
In our computation, both the UV renormalisation procedure and the IR subtraction of divergences have been realized fixing for convenience $\mu=\mu_R$, resulting in an expression for $\delta Z_e$ completely independent of any arbitrary scale.

\subsection{Fermion mass and wave-function on-shell renormalisation}
The fermion fields and masses are renormalised up to $\mathcal{O}(\alpha^2)$ with the wave-function renormalisation constants: 
\begin{align}
    &m_{f0} = \ m_f+\delta m_f =\  m_f+  \delta m_f^{(1)} + \mathcal{O}(\alpha^2) \; ,\\ \nonumber
    &f_0 = \ (Z_f)^{\frac{1}{2}} f = \left[1+\frac{1}{2}\delta Z_f^{(1)}+\frac{1}{2}\delta Z_f^{(2)} -\frac{1}{8} \left(\delta Z_f^{(1)}\right)^2 + \mathcal{O}(\alpha^3) \right] f \; .
\end{align}
The counterterm expressions has been computed with a semi-analytical evaluation fixing the on-shell renormalisation conditions satisfied from the fermion two-point vertex function. Following the same notation of \cite{Denner:1994xt,Denner:2019vbn} with $\Sigma^{f\bar{f}}_L=\Sigma^{f\bar{f}}_R=\Sigma^{f\bar{f}}$, we have:
\begin{align} \label{1st order CTS}
&\delta m_{f}^{(1)} = \ m_{f}\bigg[\Sigma^{f\bar{f}(1)}(m_{f}^{2})+\Sigma^{f\bar{f}(1)}_{S}({m_{f}}^{2})\bigg] \; ,\\ \nonumber
&\delta Z_f^{(1)} = -\Sigma^{f\bar{f}(1)}(m_{f}^{2}) - 2 m_{f}^{2}\dfrac{\partial}{\partial p^{2}} \bigg[\Sigma^{f\bar{f}(1)}(p^{2})  + \Sigma^{f\bar{f}(1)}_{S}(p^{2})\bigg]\bigg|_{p^{2}=m_{f}^{2}} \; ,\\ \nonumber
&\delta Z_f^{(2)} = -\Sigma^{f\bar{f}(2)}(m_{f}^{2}) - 2 m_{f}^{2}\dfrac{\partial}{\partial p^{2}} \bigg[\Sigma^{f\bar{f}(2)}(p^{2}) + \Sigma^{f\bar{f}(2)}_{S}(p^{2})\bigg]\bigg|_{p^{2}=m_{f}^{2}} \; .
\end{align}
Our results for the fermion counterterms are in agreement with the analytic formula present in \cite{Bonciani:2021okt} after including the self-energy dependence on the different internal masses.

\section{Photon vacuum polarization}
\label{app:vp}
The change of renormalisation scheme for the QED coupling can be expressed with a finite renormalisation between the on-shell one and $\overline{MS}$ schemes. Following the approach described in ~\cite{Degrassi:2003rw}, we define the photon vacuum polarization function from the self-energy expression:
\begin{equation}
    \Sigma^{AA}_T(q^2) = - \left(4\pi \alpha \right)  q^2  \Pi_{\gamma\gamma}(q^2) \; .
\end{equation}
Indicating with $\bar{\Pi}^{(i)}(0)$ only the $\epsilon$-divergent contribution of $\Pi^{(i)}(0)$ we therefore link the photon vacuum polarization function with our definitions of wave-function counterterms:
\begin{align}
    &\delta Z_{AA}^{OS} = \ 4\pi \alpha \ \Pi_{\gamma\gamma}(0) \implies \delta Z_{AA}^{OS(i)}= \ 4\pi \alpha^{i}\  \Pi_{\gamma\gamma}^{(i)}(0) \; ,\\\nonumber
    &\delta Z_{AA}^{\overline{MS}} = \ 4\pi \alpha_{\overline{MS}}\ \bar{\Pi}_{\gamma\gamma}(0) \implies \delta Z_{AA}^{\overline{MS}(i)}= \ 4\pi (\alpha_{\overline{MS}})^{i}\ S_\epsilon^i \ \bar{\Pi}_{\gamma\gamma}^{(i)}(0) \; .
\end{align}
The two expressions are related by the definition of the bare coupling
\begin{equation}
    \alpha_0 = \ \alpha Z_e^2 = \ \alpha  \left(Z_{AA}^{OS}\right)^{-1} = \ \alpha_{\overline{MS}}  \left(Z_{AA}^{\overline{MS}}\right)^{-1}\; ,
\end{equation} 
that we can expand up to the second order in $\alpha$ and/or $\alpha_{\overline{MS}}$ (since $\mathcal{O}(\alpha)=\mathcal{O}(\alpha_{\overline{MS}})$):
\begin{align} \label{eq:expanded_alfa-alfams_rel}
    \alpha_{\overline{MS}} =& \ \alpha -4\pi\alpha\alpha_{\overline{MS}}\left(  \Pi_{\gamma\gamma}^{(1)}(0)- \bar{\Pi}_{\gamma\gamma}^{(1)}(0) \right)  \\ \nonumber
    & - 4\pi\alpha(\alpha_{\overline{MS}})^2\left( \frac{\alpha}{\alpha_{\overline{MS}}} \Pi_{\gamma\gamma}^{(2)}(0) - \bar{\Pi}_{\gamma\gamma}^{(2)}(0)\right) + \mathcal{O}(\alpha^4)  \\ \nonumber
    =& \ \alpha \left( 1-4\pi\alpha_{\overline{MS}} \hat{\Pi}_{\gamma\gamma}^{(1)}(0) - 4\pi (\alpha_{\overline{MS}})^2 \hat{\Pi}_{\gamma\gamma}^{(2)}(0)  + \mathcal{O}(\alpha^3) \ \right) \; ,
\end{align}
where the factors $S_\epsilon^i$ have been removed by rescaling the t'Hooft parameter in $\Pi_{\gamma\gamma}(0)$ with $\mu\rightarrow \sqrt{e^{\gamma_E}/4\pi}\ \mu$. In the last row we exploited the fact that $\alpha/\alpha_{\overline{MS}}=1+O(\alpha)$ and we defined $\hat{\Pi}_{\gamma\gamma}^{(i)}(0)$ as the finite remainder from the difference $\Pi_{\gamma\gamma}^{(i)}(0) - \bar{\Pi}_{\gamma\gamma}^{(i)}(0)$.
Equation \ref{eq:expanded_alfa-alfams_rel} directly leads to the definition of $\Delta \alpha_{\overline{MS}}(m_Z^2)$ after setting $\mu_R=m_Z$.
The photon vacuum polarisation can be decomposed as:
\begin{align} \label{eq:Pigammagamma_decomposition}
     \Pi_{\gamma \gamma} (0) = \ &\Pi_{\gamma \gamma}^{(\text{lep})}(0) +\Pi_{\gamma \gamma}^{(\text{top})} (0) +\Pi_{\gamma \gamma}^{(5)}  (0) \\ \nonumber
     =\ &\Pi_{\gamma \gamma}^{(\text{lep})}(0) +\Pi_{\gamma \gamma}^{(\text{top})}(0) +\text{Re}(\Pi_{\gamma \gamma}^{(5)} (m_Z^2)) - \frac{\Delta \alpha_{\text{had}}(m_Z^2)}{4\pi\alpha}  \; ,
\end{align}
where $(\text{lep})$ and $(\text{top})$ are the perturbative contributions from the leptons and the massive top quarks, $(5)$ is the contribution from the five light quarks and $\Delta \alpha_{\text{had}}(m_Z^2) = 4\pi\alpha\left(\text{Re}(\Pi_{\gamma \gamma}^{(5)} (m_Z^2))-\Pi_{\gamma \gamma}^{(5)}  (0)\right)$ is derived via a dispersion relation from the experimental cross section of $e^+ e^- \rightarrow \text{hadrons}$.
In conclusion we obtain:
\begin{align}
     \Delta \alpha_{\overline{MS}}(m_Z^2) = & \ \Delta \alpha^{(1)}_{\overline{MS}}(m_Z^2)  + \Delta \alpha^{(2)}_{\overline{MS}}(m_Z^2) + \mathcal{O}(\alpha^3)= \\ \nonumber
     = &\ - 4\pi \alpha_{\overline{MS}} \left[\hat{\Pi}_{\gamma\gamma}^{(\text{lep})(1)}(0) + \hat{\Pi}_{\gamma\gamma}^{(\text{top})(1)}(0) +\text{Re}(\hat\Pi_{\gamma \gamma}^{(5)(1)} (m_Z^2)) - \frac{\Delta \alpha_{\text{had}}(m_Z^2)}{4\pi\alpha} \right] \\ \nonumber
     & \ - 4\pi (\alpha_{\overline{MS}})^2 \left[\hat{\Pi}_{\gamma\gamma}^{(\text{lep})(2)}(0) + \hat{\Pi}_{\gamma\gamma}^{(\text{top})(2)}(0) +\text{Re}(\hat\Pi_{\gamma \gamma}^{(5)(2)} (m_Z^2)) \right]+ \mathcal{O}(\alpha^3)  \; .
\end{align}
The term $\Delta\alpha_{\text{had}}(m_Z^2)$ starts at $\mathcal{O}(\alpha)$ and takes part to the definition of $\Delta \alpha^{(1)}_{\overline{MS}}(m_Z^2)$, thus affecting the change of scheme for the virtual amplitude starting from NLO.

\bibliography{long}
\bibliographystyle{JHEP}

\end{document}